\documentclass[10pt,letterpaper,twocolumn]{article} 

\usepackage{ol2}
\usepackage[draft]{hyperref}
\usepackage{subfigure}
\usepackage{amsmath,amssymb,graphicx}

\newcommand{\Pt}{$P_\textrm{th}$ }

\begin{document}

\twocolumn[ 

\title{Thresholdless discrete surface solitons and stability switchings in periodically curved waveguides}


\author{H.\ Jiang,$^1$ H.\ Susanto,$^{1,*}$ T.M.\ Benson,$^2$ and K.A.\ Cliffe$^{1}$}
\address{
$^1$School of Mathematical Sciences, University of Nottingham, University Park, Nottingham, NG7 2RD, UK
\\
$^2$Electrical Systems and Optics Division, 
University of Nottingham, University Park, Nottingham, NG7 2RD, UK\\
$^*$Corresponding author: hadi.susanto@nottingham.ac.uk
}

\begin{abstract}
We study numerically a parametrically driven discrete nonlinear Schr\"odinger equation modelling periodically curved waveguide arrays. We show that discrete surface solitons persist, but their threshold power is altered by the drive. There are critical drives at which the threshold values vanish. We also show that parametric drives can create resonance with a phonon making a new barrier for discrete solitons. By calculating the corresponding Floquet multipliers, we find that the stability of symmetric and antisymmetric off-side discrete surface solitons switches approximately at the critical drives for thresholdless solitons.
\end{abstract}


] 

Parametric drives have been proposed as a means to control localized waves in a linear Schr\"odinger system, which in the undriven case would simply disperse \cite{dunl86}. Such drives are created among others in periodically curved waveguide arrays \cite{lenz03,long05}. The dynamic localization has been confirmed experimentally in \cite{long06,szam09}. The same dynamic control has been applied in Bose-Einstein condensates to suppress quantum tunneling of particles trapped in a potential well by shaking the potential \cite{lign07,sias08,zene09} (see also the review \cite{mors10}). Periodically curved waveguides have also been used to create defect-free surface modes \cite{szam08a,gara08}.

Note that the above results are in the linear regime. When nonlinearity is present, fundamental properties of dynamic localization will be altered. It is indeed the case with the so-called coherent destruction of tunneling \cite{gros91,gros91b}, i.e.\ dynamic localization in a finite domain. The parameter values for tunneling suppression, which are isolated points in the linear systems, become finite intervals for the nonlinear counterparts \cite{luo07} (see also \cite{jian12}). The method can still be proposed for dispersion management of, e.g., a nonlinear atomic wavepacket \cite{cref09} and gap-solitons \cite{blud04}.

Discrete spatial solitons occur when the nonlinearity effect balances the dispersion and diffraction of the system. In periodically curved optical waveguides they were reported experimentally in \cite{szam08}. Different from self-trapping in straight waveguides, the solitons are obtained after transitional self-induced beam broadening. Nonlinear surface waves were observed experimentally in \cite{qi09}. See, e.g., \cite{gara12} for a review.

In this letter, we show that the parametric drives decrease the threshold power \Pt of discrete surface solitons reported in \cite{makr05,sunt06,moli06}. More importantly, there are critical amplitudes where the thresholds vanish. Nevertheless, curved waveguides can be parasitic when the discrete localisation is in resonance with the drive. Finally, we show that symmetric and antisymmetric off-side, i.e.\ intersite and twisted, modes switch stability at approximately the threshold driving amplitudes for thresholdless discrete surface solitons.

The optical field $a_n$ along the propagation direction $z$ is modelled by \cite{long05}
    \begin{equation}
    i\dot{a}_n = c[e^{-i\dot{\tilde{x}}_0(z)}a_{n+1}+e^{i\dot{\tilde{x}}_0(z)}a_{n-1}]+\delta\left|a_n\right|^2a_n-q a_n,
    \label{gov}
    \end{equation}
where $c>0$ and $\delta>0$ are the waveguide coupling and the nonlinearity coefficient, respectively. The defocusing case $\delta<0$ can obtained using the transformation $a_n\to(-1)^na_n$. The drive is $\dot{\tilde{x}}_0(z)=(n_s\alpha /h)\dot{x}_0(z)$, where $\alpha$ is the separation distance between the waveguides, $n_s$ is the substrate refractive index, $x_0(z)$ is the physical modulation profile, and $h$ is the inverse of the light wavenumber. Here, we take $\tilde{x}_0=-A\,\cos(\omega z)$. By proper scaling, one can take $\delta=\omega=1$. The propagation constant $q$ is a control parameter related to the constant field power $P=\sum_n|a_n|^2$. Numerically its presence removes the commensurability requirement between the oscillation frequency of the wavefield and that of the drive. Eq.\ (\ref{gov}) is solved for periodic orbits (see, e.g., \cite{morg02} for the methods). The stability of a periodic orbit is examined by calculating its Floquet multipliers. 

\begin{figure}[tbh]
\centering
\subfigure[]{\includegraphics[width=4.4cm]{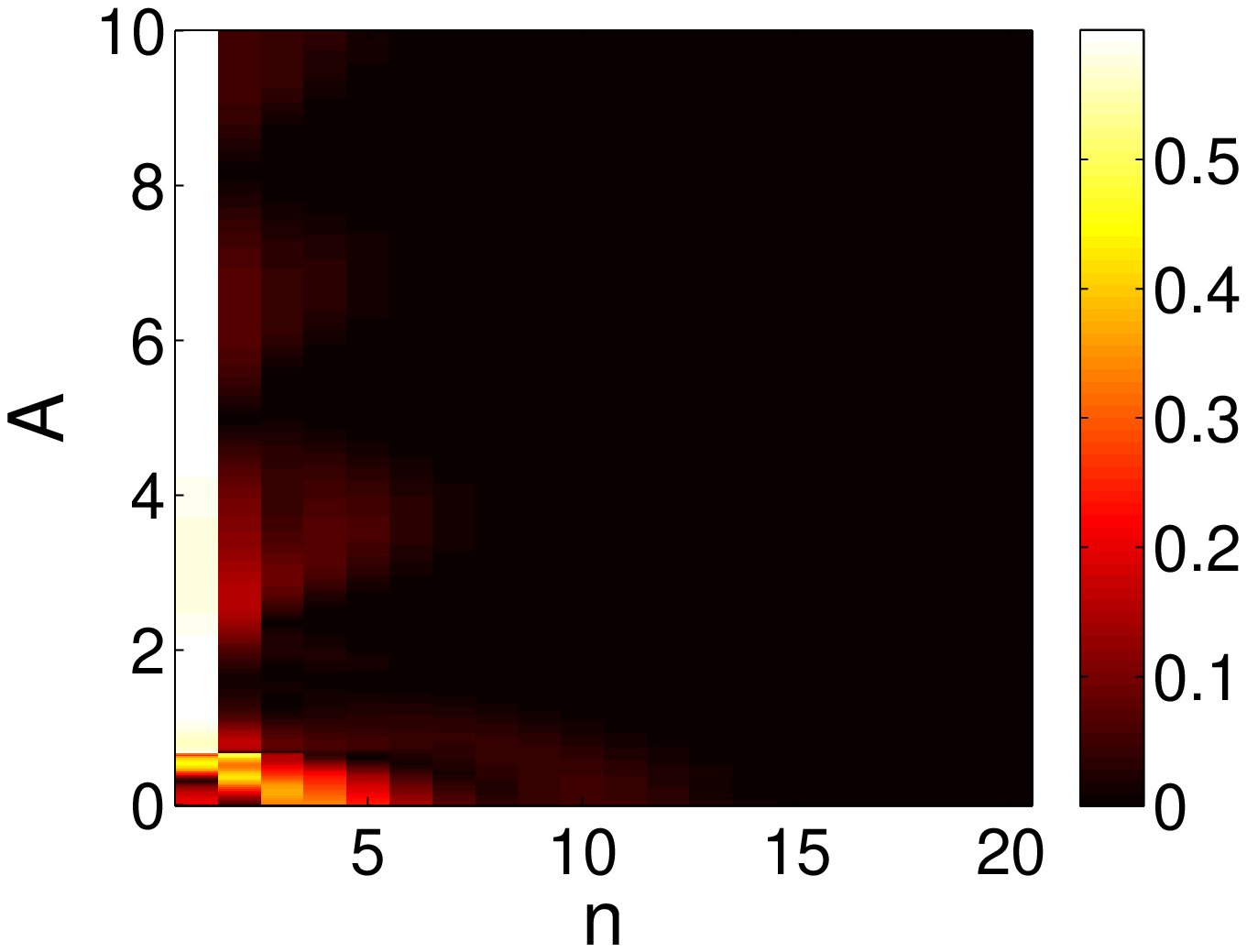}}%
\subfigure[]{\includegraphics[width=4.4cm]{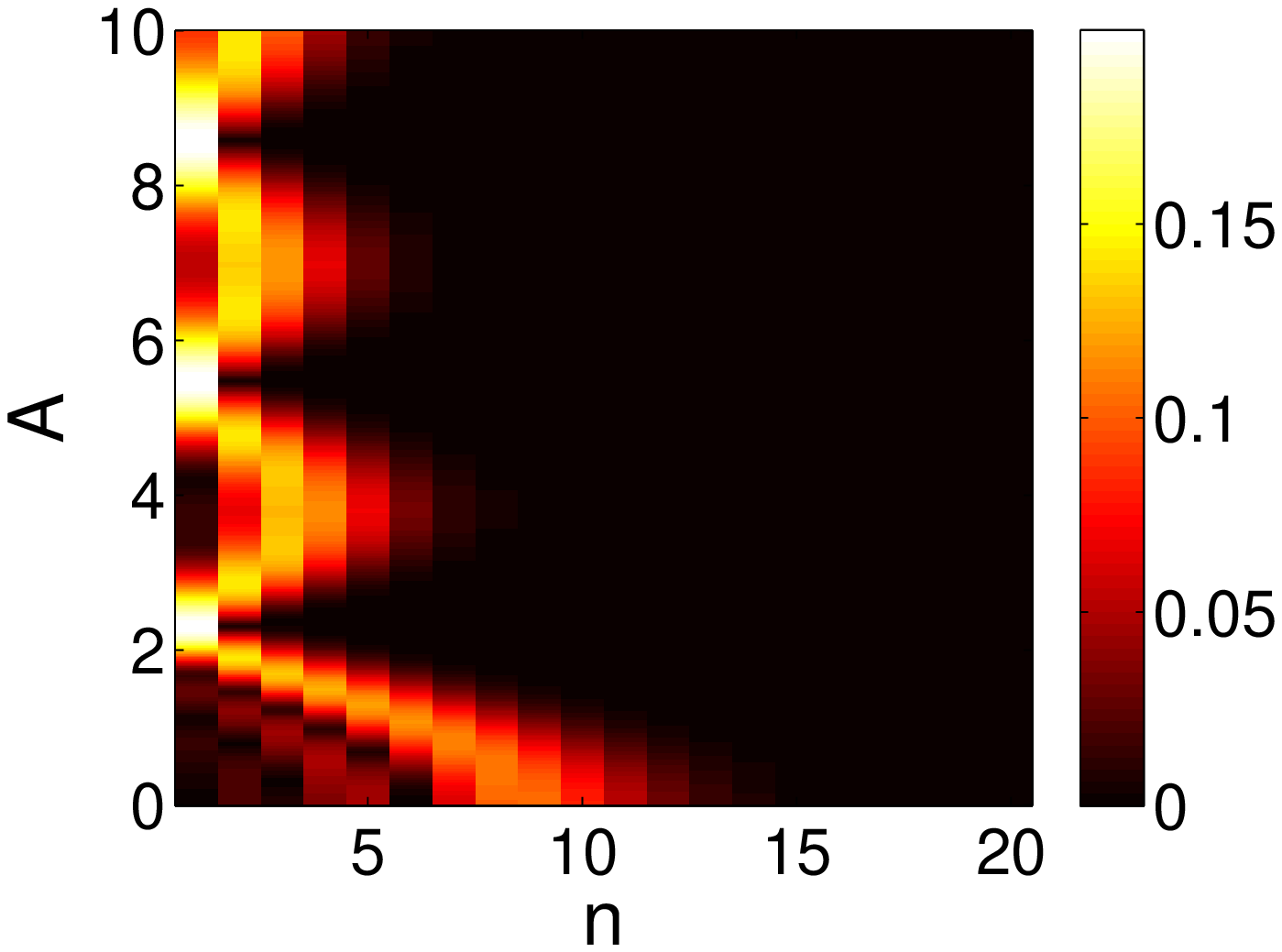}}
\caption{Field distributions $|a_n|$ at $z=50$ mm for varying $A$ with $c=0.1$. Initially only the most left waveguide is excited with (a) $a_1(0)=0.6$, (b) $a_1(0)=0.2$.}
\label{fig1}
\end{figure}

To illustrate the effect of curvature on the formation of discrete surface solitons, we plot in Fig.\ \ref{fig1}(a,b) the output intensity $|a_n(50)|$ for the initial condition $a_{n\neq1}(0)=0$, $a_1(0)=0.6$ and $0.2$, respectively. These correspond to the case of above and below \Pt of the undriven case, i.e.\ $P\approx0.33$ (calculated using the method of \cite{makr05,moli06}). In the first case, when $A=0$ a discrete soliton does not form because the repulsive force of the surface \cite{moli06} is large enough to push the excitation to the right. Nevertheless, a relatively small driving amplitude $A\approx0.67$ is enough to form a surface soliton. More interestingly, for the second case we observe the formation of discrete solitons at particular values of $A$ even though the power is far below \Pt of the undriven case, which will be explained by studying waves of (\ref{gov}) that are periodic-in-$z$ but localized in the transverse direction.

\begin{figure}[tbh]
\centering
\subfigure[]{\includegraphics[width=4.4cm]{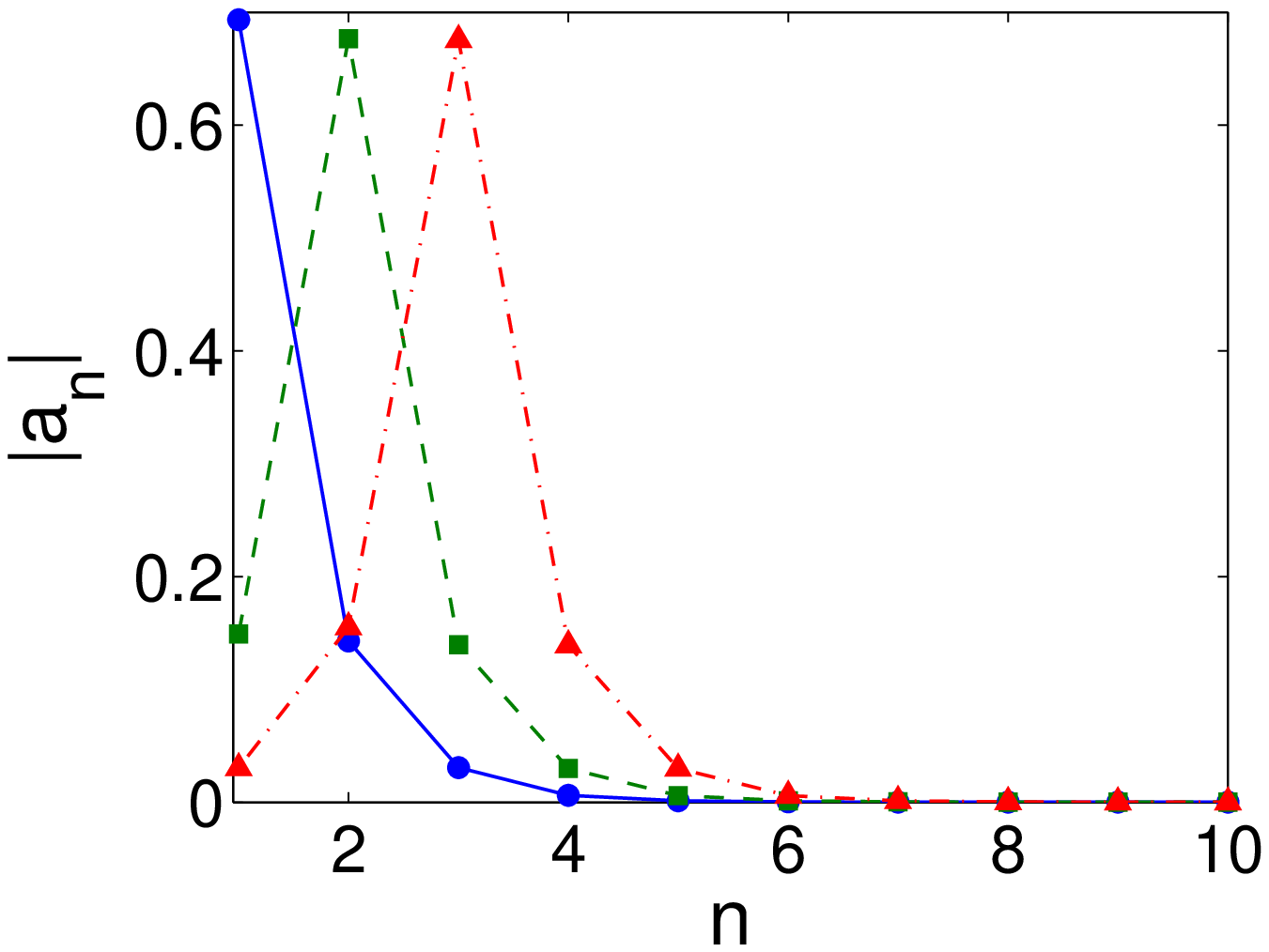}}%
\subfigure[]{\includegraphics[width=4.4cm]{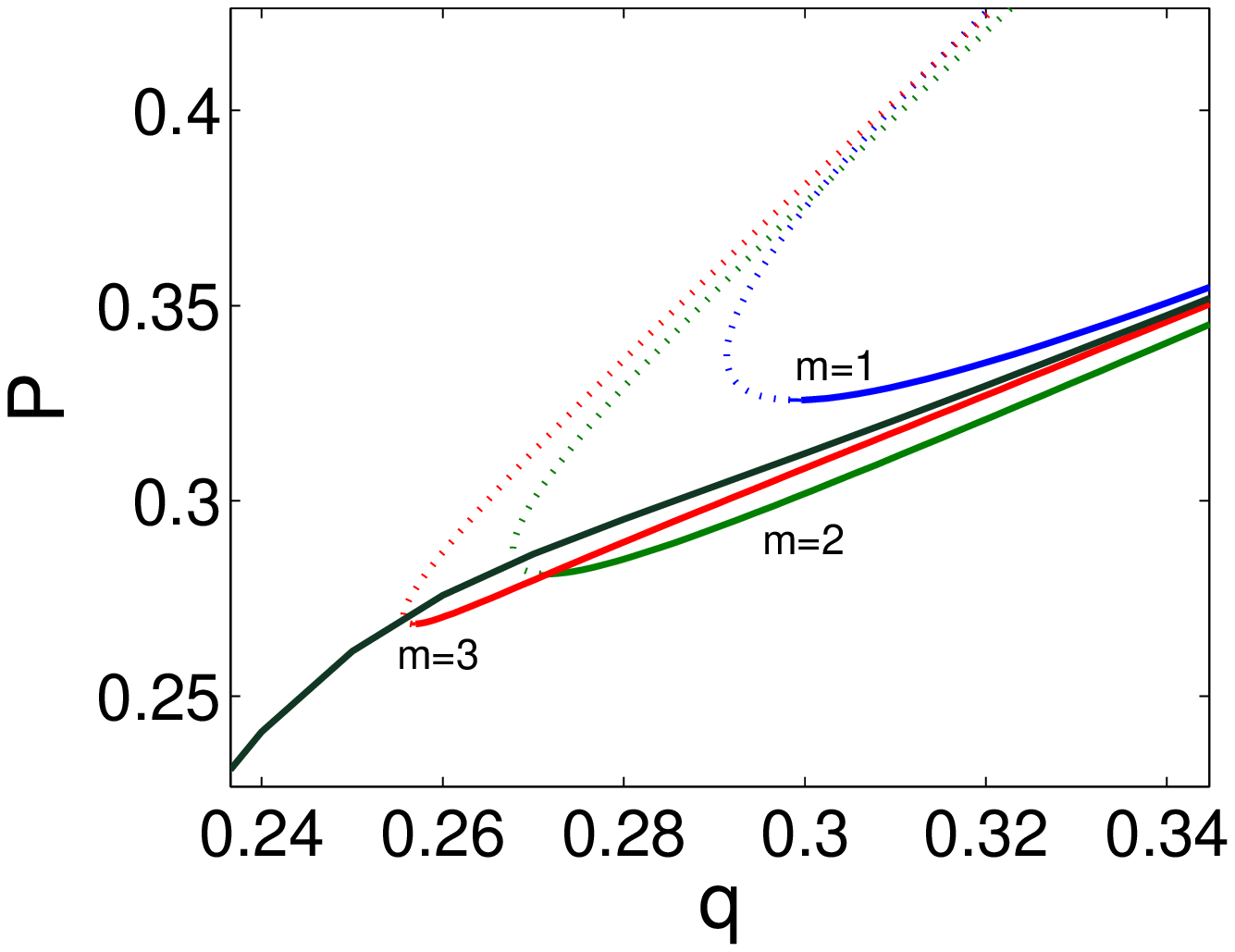}}
\subfigure[]{\includegraphics[width=4.4cm]{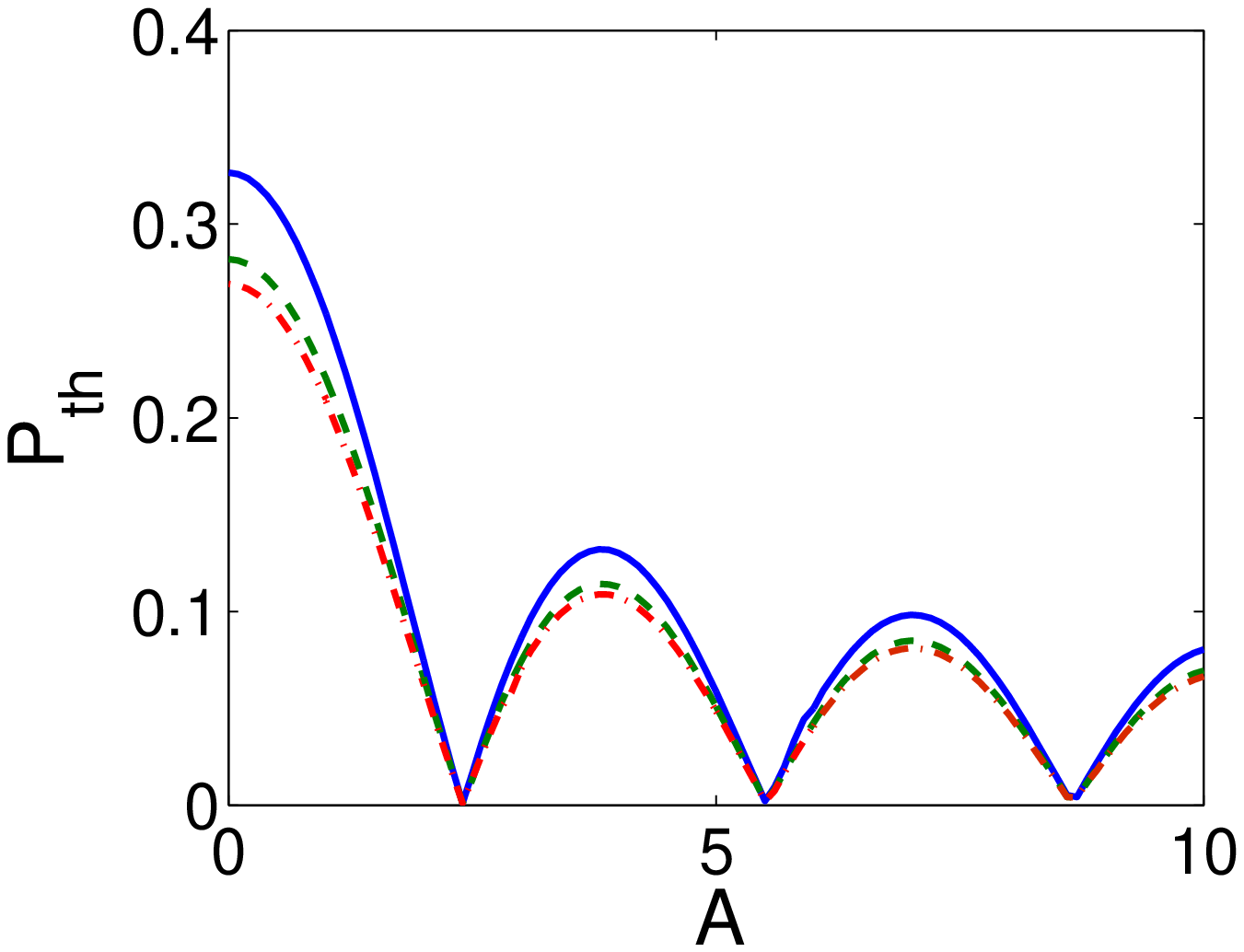}}%
\subfigure[]{\includegraphics[width=4.4cm]{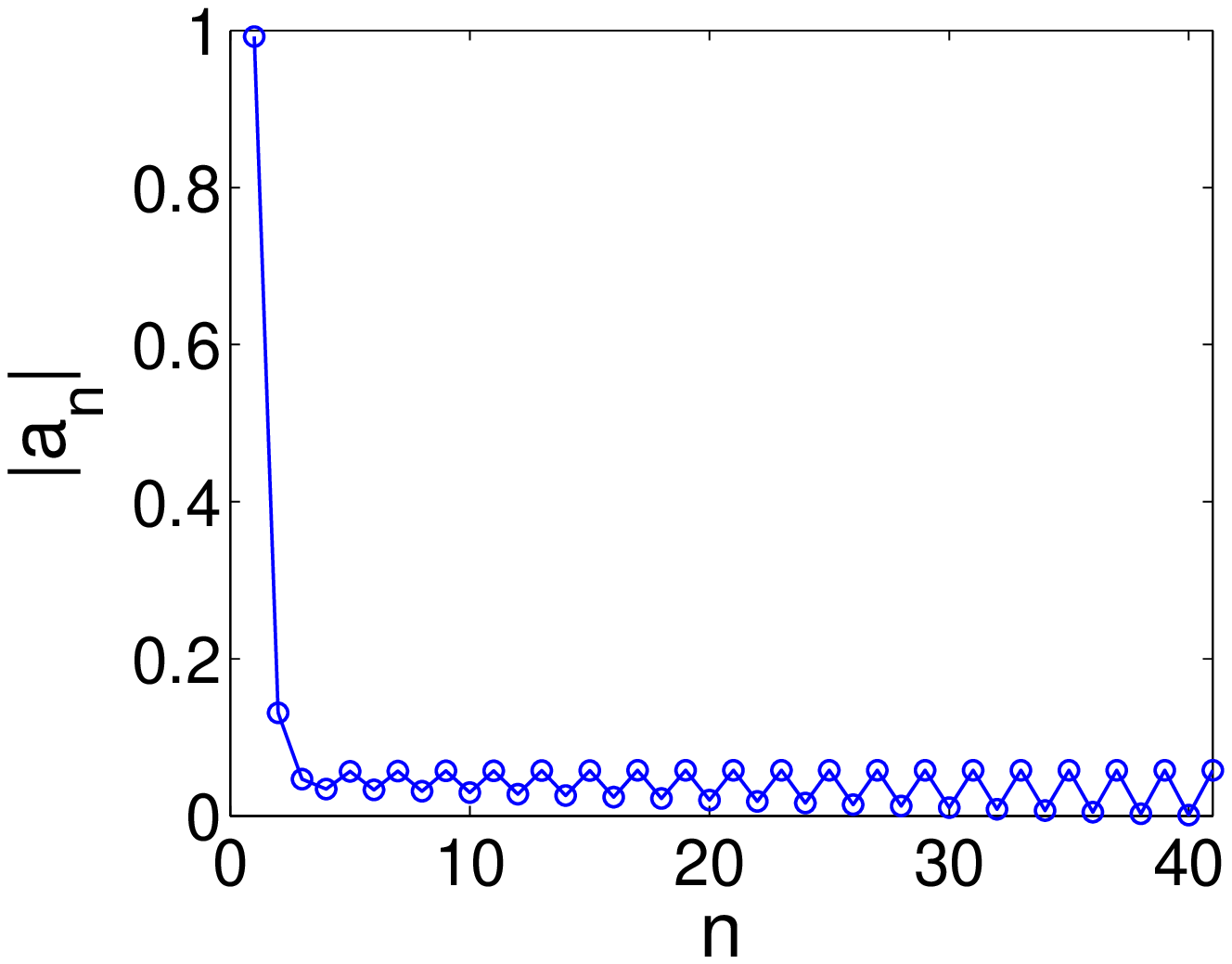}}
\caption{\label{Fig1} (a) Amplitude profile at $z=0$ for discrete surface solitons centered at different sites near the surface with $c = A=0.1$ and $q = 0.5$. 
(b) The power of discrete solitons in (a) for varying $q$. The solid black line corresponds to discrete 'bulk' solitons. Dotted curves show instability. (c) \Pt as a function of $A$. (d) A surface phonobreather for the same parameter values and $q=1$.}
\end{figure}

We depict numerically exact discrete surface solitons of \eqref{gov} centered at the $m$th waveguide in Fig.\,\ref{Fig1}(a) for $c=0.1,\,A=0.1$, and $q=0.5$. 
In the presence of parametric drives, we show in Fig.\,\ref{Fig1}(b) the power of discrete surface solitons for varying $q$. One can observe that as $q$ decreases the power also decreases. Yet, there is a critical value of $q$ where the power starts increasing. This implies that there is a minimum power for surface modes to exist, as for the undriven case. As a comparison, we also present as solid black line the power variation for discrete 'bulk' solitons that have no threshold power, i.e.\ \Pt$=0$. We also calculate the stability of the modes where there is a pair of Floquet multipliers at $+1$ at $P_\textrm{th}$. The multipliers leave the unit circle as the continuation passes the minimum power point, i.e.\ the discrete surface solitons are unstable and represented by the dotted lines in Fig.\,\ref{Fig1}(b).

Studying \Pt further, we found that it is a function of $A$. Performing similar continuations as in Fig.\ \ref{Fig1}(b) for various driving amplitudes, \Pt is plotted in Fig.\,\ref{Fig1}(c). One can observe that \Pt decreases with the increase of $A$. The decrease in \Pt is responsible for the formation of surface solitons observed for nonvanishing $A$ in Fig.\ \ref{fig1}(a). More importantly, in Fig.\ \ref{Fig1}(c) there are critical amplitudes where \Pt$=0$. At these values of $A$, we obtain discrete surface solitons without any threshold power. The formation of discrete solitons in Fig.\ \ref{fig1}(b) occurs at these particular driving amplitudes.

Despite the constructive effect, we also observed that the parametric drive can be parasitic by inducing radiation from the excited site. In that case, we could not obtain genuinely localized waves and instead found discrete surface solitons with non-zero tails as shown, e.g., in Fig.\ \ref{Fig1}(d). Such solutions are commonly referred to as phonobreathers, i.e.\ localised waves that are in resonance with a phonon (see, e.g., \cite{morg02} and references therein). As the phonon band of (\ref{gov}) is the interval $\pm [q-2c, q+2c]$, it is straightforward to obtain the resonance condition 
$    k \in [q-2c, q+2c]$,
for an arbitrary integer $k$. Thus, discrete solitons cannot exist within these regions.

\begin{figure}[tbh]
\centering
\subfigure[]{\includegraphics[width=4.4cm]{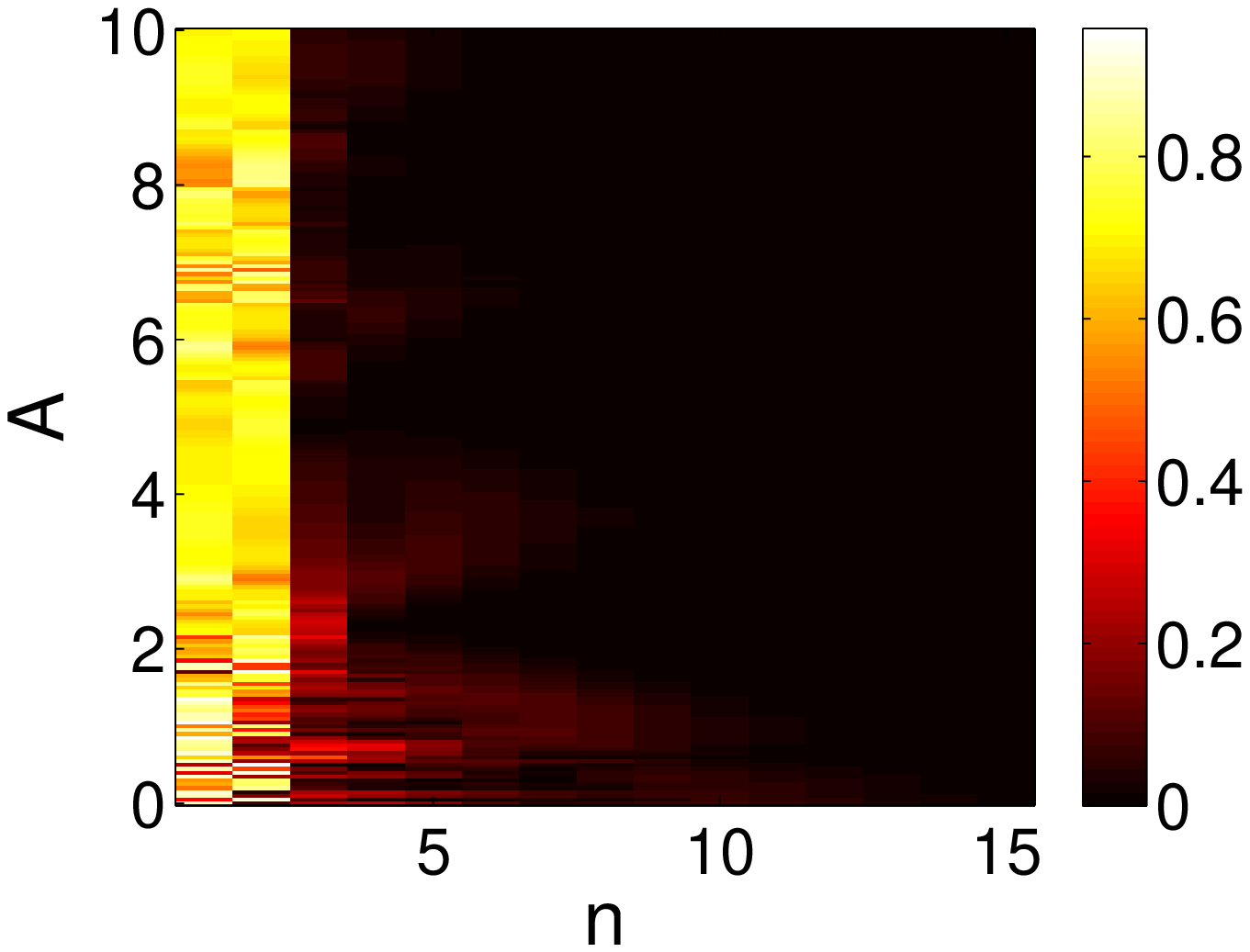}}%
\subfigure[]{\includegraphics[width=4.4cm]{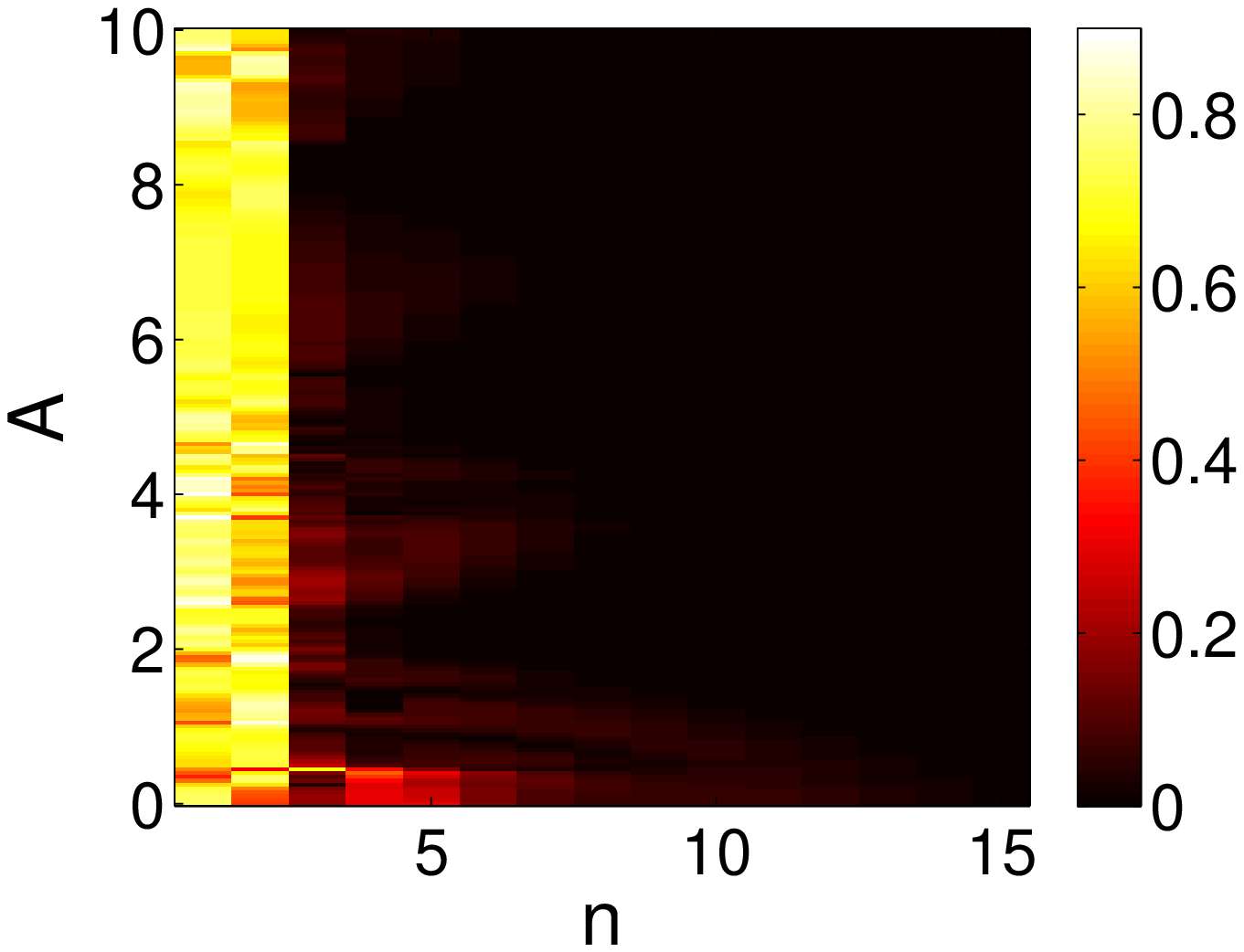}}
\subfigure[]{\includegraphics[width=4.4cm]{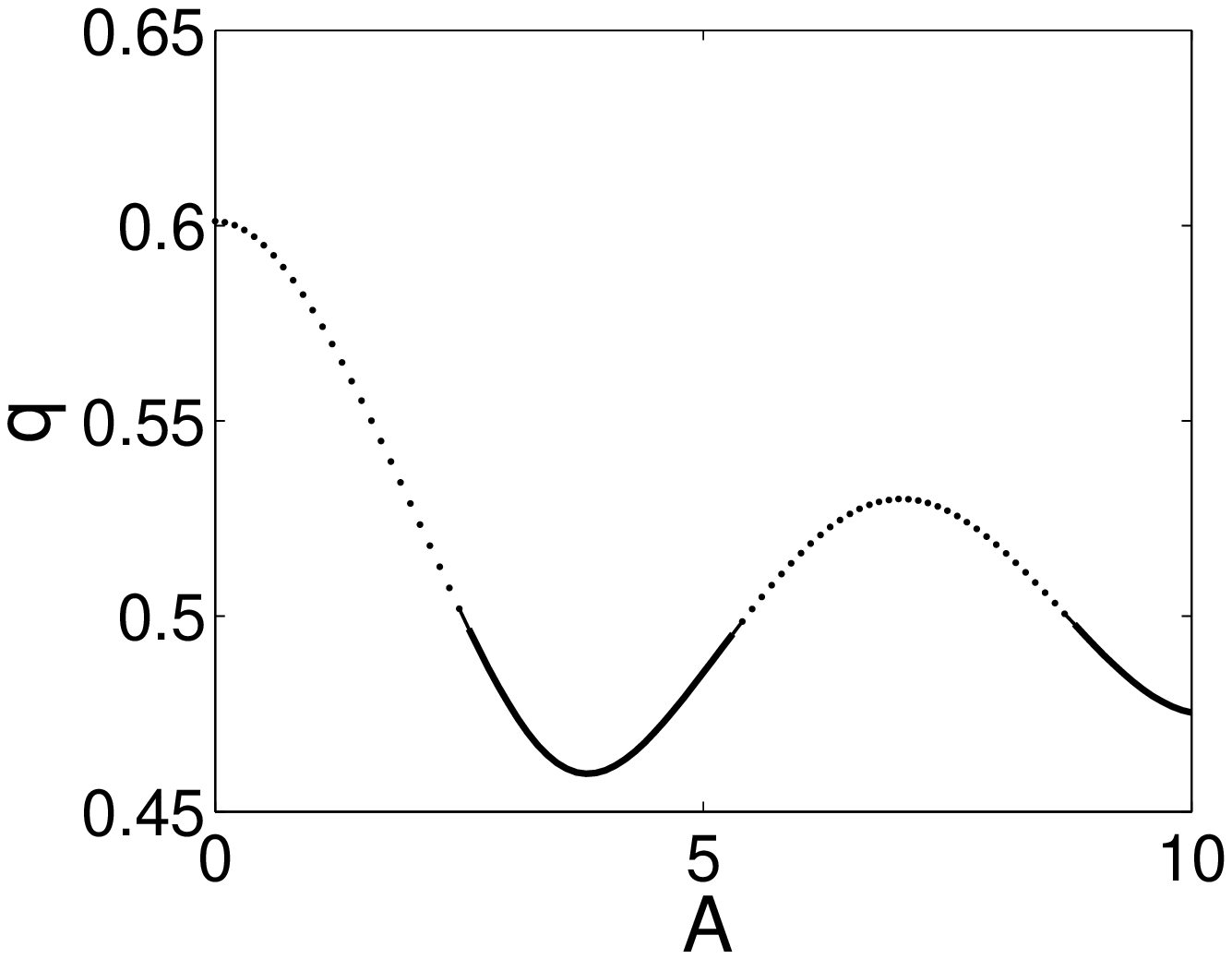}}%
\subfigure[]{\includegraphics[width=4.4cm]{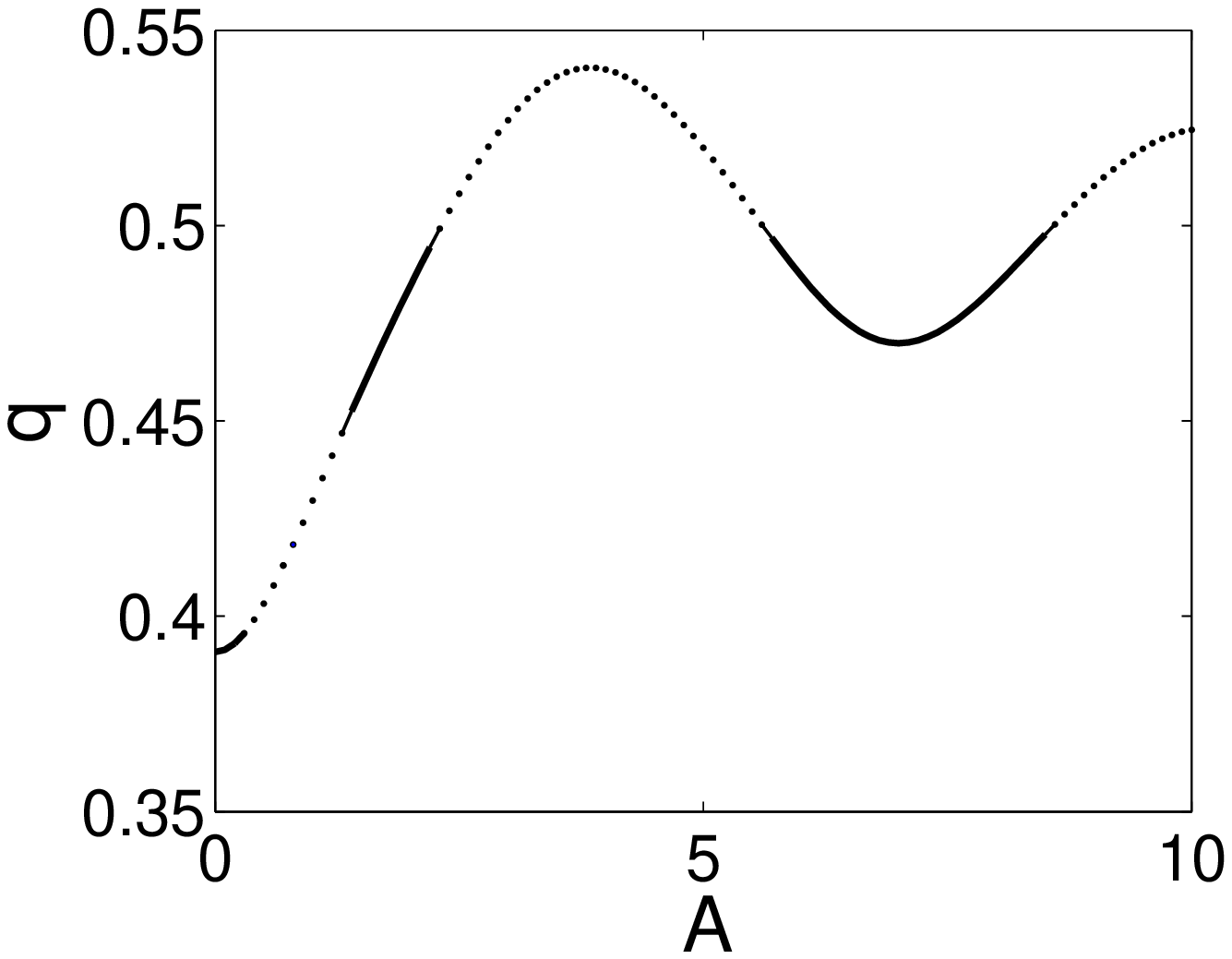}}
\caption{\label{fig3} (a-b) The same as Fig.\ \ref{fig1}, but for initial conditions $a_{n\neq1,2}=0$, $a_{1,2}=\sqrt{0.5}$ (a) and $a_{1}=-a_2=\sqrt{0.5}$ (b). (c-d) The stability diagram of the intersite (c) and twisted (d) mode for fixed power $P=1$. Shown is $q$ against $A$. Dotted lines correspond to unstable solutions.}
\end{figure}

We have also studied symmetric and antisymmetric two-excited site modes, which are referred to as (in-phase) intersite and (out-of-phase) twisted modes. For discrete 'bulk' solitons in infinite arrays it is known that in-phase modes are always unstable and out-of-phase modes have a stability interval. Figs.\ \ref{fig3}(a-b) show the field distributions at $z=50$ of the initial state where only the first two left waveguides are excited in-phase and out-of-phase, respectively. We observe that there are some intervals of $A$ where intersite discrete surface solitons are obtained. On the other hand, there are also intervals where twisted surface solitons are not obtained, even though the results are less clear than the first ones. These indicate that periodically curved waveguides can support in-phase discrete solitons as well as destroy out-of-phase modes.

We have also sought the exact periodic intersite and twisted modes of (\ref{gov}). The upper (dotted) curves in Fig.\ \ref{Fig1}(b) correspond to intersite surface solitons. We depict in Fig.\ \ref{fig3}(c-d) the stability of those modes as a function of $A$. The power is fixed and the same as the initial conditions. We observe that the drives can indeed stabilize and destabilize unstable intersite and stable twisted states, respectively. The stability regions of the in-phase solitons are in agreement with the region for the formation of two-excited site surface solitons in Fig.\ \ref{fig3}(a) with the stability switchings occuring approximately at the critical drives for thresholdless discrete surface solitons (see Fig.\ \ref{Fig1}(c)). As for twisted modes, it is rather only the stability window of Fig.\ \ref{fig3}(d) for $A>5$ that is comparable with that of Fig.\ \ref{fig3}(b). While all the stability switchings are due to exponential instability, i.e.\ Floquet multipliers leave at $+1$, we observe that the first stability switch in Fig.\ \ref{fig3}(d) as $A$ increases from 0 is  
due to a Hamiltonian Hopf bifurcation. 

Analysis of (\ref{gov}) is usually performed through its averaged equation \cite{long05}
    \begin{equation}
    i\dot{a}_n = cJ_0[a_{n+1}+a_{n-1}]+\delta\left|a_n\right|^2a_n-q a_n,
    \label{av}
    \end{equation}
where $J_0(A)$ is a Bessel function of the first-kind. Thus, $A\neq0$ decreases the effective coupling between the waveguides yielding smaller \Pt than the undriven case. Therefore, thesholdless discrete surface solitons are expected to be achieved when $J_0(A)=0$. Despite the similarity with the condition for dynamic localizations, the resonance here is related to self-trapped states, i.e.\ no diffusion and dispersion of fields. As for in-phase off-side modes, using (\ref{av}) one would expect the stability switching to occur when $J_0<0$, i.e.\ effectively the modes become twisted modes. Using a similar observation, the twisted mode should have been unstable when $J_0(A)>0$. Nevertheless, \eqref{av} could not predict the first instability window in Fig.\ \ref{fig3}(d) due to a quartet of multipliers. \eqref{av} is expected to be valid when $e^{\pm i\dot{\tilde{x}}_0}$ rapidly oscillates, i.e.\ $A\omega\gg1$ when all the other parameters are of $\mathcal{O}(1)$. 

To conclude, we have shown numerically that periodically curved waveguides can control the formation as well as annihilation of discrete (surface or 'bulk') solitons. 
The parametric drives can also be parasitic to discrete solitons by creating resonances with the phonon.

HJ, HS, and TMB acknowledge the partial financial support of a University of Nottingham Interdisciplinary High Performance Computing (iHPC).


\end{document}